\documentclass[prb,preprint]{revtex4-1}
\usepackage{amsmath}
\usepackage{graphicx}
\usepackage{epsfig}

\begin{document}

\title{Direction dependency of extraordinary refraction index in uniaxial
nematic liquid crystals}
\author{Jerneja Pavlin}
\email{jerneja.pavlin@pef.uni-lj.si}
\author{Nata\v{s}a Vaupoti\v{c}}
\author{Mojca \v{C}epi\v{c}}
\date{\today}

\begin{abstract}
The article presents a straightforward experiment that directly and
illustratively demonstrates double refraction. For this purpose, two liquid
crystalline cells were designed, which enable qualitative and quantitative
measurements of the extraordinary refractive index direction dependency in a
uniaxial nematic liquid crystal.
\end{abstract}

\maketitle


%\altaffiliation[Also at ]{home.} % optional
\affiliation{University of Ljubljana, Faculty of Education, Ljubljana}

%optional

\affiliation{University of Maribor, Faculty of Natural Sciences and Mathematics, Maribor\\
and Jo\v{z}ef Stefan Institute, Ljubljana}

\affiliation{University of Ljubljana, Faculty of Education, Ljubljana\\
and Jo\v{z}ef Stefan Institute, Ljubljana}

\section{Teaching the concept of anisotropy}

When a physical property in the material varies with the direction, the
material is said to be anisotropic. Anisotropy of physical properties is a
crucial property of materials that are nowadays extremely important in
science and technology. An example of such materials involves the liquid
crystals that enable production of several devices used in everyday life
(flat-screen TVs, notebooks, iPads, calculators, etc.). Even though students
are in touch with such devices on a daily basis, they are not aware of
anisotropy as a key property of the materials that enable the hi-tech
products they use so eagerly. This ignorance might be due to the fact that
the concept of anisotropy and in particular its consequences are rather
difficult to comprehend. \cite{nclc, knitting} It is thus important to
introduce the concept of anisotropy as soon as possible, taking into account
students' knowledge of physics and mathematics. \cite{ciferno, Pavlin,
pieranski}

In this paper we focus on birefringence as one of the most widely used
properties of anisotropic materials. In a birefringent material there are
two distinct waves that can propagate in a general direction determined by
the wave vector direction ($\vec{k}$). The two waves are linearly polarized,
their polarizations being perpendicular. They travel with different phase
velocities, which means that the indices of refraction are different for the
two waves although their wave vectors are in the same direction. Both
refractive indices depend on the direction of $\vec{k}$.

An important property of light propagation in anisotropic materials is that
the energy (defined by the Poynting vector $\vec{S}=\vec{E}\times \vec{H}$,
where $\vec{E}$ is the electric and $\vec{H}$ the magnetic field) does not
propagate in the same direction as the wave vector (i.e., the wave fronts
propagate in a different direction than the energy). The ray that we see is
the direction of energy propagation. So, in order to avoid confusion, one
should strictly state which velocity is considered (phase or ray velocity),
and when considering the direction of light propagation, state whether the
ray or wave vector direction is being considered.

In a general anisotropic medium there are two directions of wave vector
propagation, along which the two waves travel with the same phase velocity.
These two directions are called the optic axes, and the material is said to
be biaxial. In the biaxial material there are also two directions, along
which the ray velocities for both waves are equal. These axes are called the
ray axes and are distinct from the optic axes. When optic axes are
identical, material is uniaxial. In uniaxial materials the optic and the ray
axes are the same. In optically uniaxial materials the phase and ray
velocities of one of the two waves are equal, independent of the direction
of propagation. This wave is called \textit{the ordinary wave}. For the
second wave, the phase and ray velocities are different in magnitude and
direction, and they both depend on the direction of propagation. This wave
is called \textit{the extraordinary wave}.

The most easily observed and striking optical property of transparent
anisotropic materials is double refraction (birefringence): the light
incident on the interface between the isotropic and anisotropic material
refracts into the anisotropic material in such a way that there are in
general two rays of light propagating through the material in different
direction; the light in both rays is polarized (even if the incident light
is not polarized); light polarizations in the two rays are perpendicular.

Double refraction is usually demonstrated by observing the doubling of a
text observed through the calcite. \cite{calcite} When a polarizer is placed
behind the calcite (or in front of it), one of the figures disappears if the
polarizer's transmission direction coincides with the polarization of the
transmitted light. Although the explanation for this phenomenon is rather
straightforward for a trained physicist, it is usually not so easily
comprehended by students. A more straightforward experiment demonstrates the
splitting of the unpolarized incident light beam into two beams of linearly
polarized light. One needs a large birefringent crystal that is thick enough
so that two transmitted beams can be observed as two light spots on a
distant screen. By changing the direction of the incident light by rotating
the crystal, one can observe the direction dependency of the extraordinary
wave refractive index in uniaxial crystals, as well as the direction
dependency of both indices in biaxial crystals. Unfortunately, the accuracy
of quantitative measurements for both indices is poor, as any, even slight
non-parallelism of the crystal surfaces results in a significant change in
the transmitted light direction. Since the most easily accessible crystals
are biaxial (e.g., quartz), this additionally complicates the comparison of
the experimental results and the theoretical considerations.

Liquid crystals are optically anisotropic materials that are easy and cheap
to obtain; even more, they can be synthesized in a school laboratory, not
only at the university level but also at the high school level. \cite%
{Liberko, Verbit, Wright, Pavlin} By using liquid crystals, the concept of
anisotropy can be qualitatively introduced even at the elementary school
level. At the high school level, anisotropic properties can be introduced
through a set of carefully designed experiments. The concept can be
efficiently reinforced at the university level, with an interdisciplinary
teaching module consisting of lectures and lab work.\cite{Pavlin, Eurasia}
The module is appropriate for both social and natural science students.
While such a teaching module has proven to be very efficient in achieving
the conceptual understanding of optical anisotropy, Physics students also
require experiments that allow quantitative measurements. Liquid crystals
offer a possibility to measure angular dependency of the refractive index in
uniaxial materials. In this paper we present an experiment that enables
quantitative measurements of the refractive index for light propagating at
angles ranging from $0^{\circ }$ to $90^{\circ }$ with respect to the optic
axis.

\section{Propagation of light in anisotropic media}

In an anisotropic dielectric material, the polarization ($\overrightarrow{P}$%
) in the material depends on the direction of the external electric field ($%
\overrightarrow{E}$), and it is, in general, not in the direction of the
external field. This property is described by the electric susceptibility (%
\underline{$\chi $}) being a tensor quantity (not scalar, as in isotropic
materials), and the relation between the material polarization and the
electric field is:%
\begin{equation}
\overrightarrow{P}=\varepsilon _{0}\underline{\chi }\overrightarrow{E}\quad .
\label{eq:pol}
\end{equation}%
Light is the propagation of an electromagnetic wave. The electric field in
the electromagnetic wave interacts with the material and induces
polarization, which varies over time with the same frequency as the
frequency of light. Interaction of light with matter reduces the speed of
light. In materials where the field induced polarization is larger
(materials with greater electric susceptibility), the speed of light is
lower. The ratio between the speed of light in a vacuum and in a particular
material is given by the index of refraction. Since in anisotropic materials
polarization in the material depends on the direction of the external
electric field, in such materials the index of refraction varies with the
direction of light propagation, and it depends on the direction of the
electric field (i.e., the direction of light polarization). We point out
that polarization of light (direction of $\overrightarrow{E}$ in light)
should not be confused with material polarization ($\overrightarrow{P}$),
which is a result of $\vec{E}$ in light. The reader should once again
carefully examine Eq. (\ref{eq:pol}) to fully grasp the difference.

Propagation of light is described by the wave equation. We shall assume a
nonmagnetic material (no magnetization due to the external magnetic field)
that contains no volume charge and no conducting current. Assuming a
monochromatic plane wave with frequency $\omega $ propagating in any
direction given by the wave vector $\overrightarrow{k}$, the electric and
magnetic field vectors may be described by the harmonic representation, e.g. 
$\overrightarrow{E}=\overrightarrow{E}_{0}e^{i(\overrightarrow{k}\cdot 
\overrightarrow{r}-\omega t)}$ . The general \textit{plane wave equation} in
the anisotropic media is \cite{nasvet-valen}
\begin{equation}
\overrightarrow{k}\times (\vec{k}\times \vec{E})+k_{0}^{2}\underline{%
\varepsilon }\vec{E}=0\quad ,  \label{eq:valovnaen}
\end{equation}%
where $k_{0}$ is the magnitude of the wave vector in a vacuum, and 
\underline{$\varepsilon $}$=\underline{I}+$\underline{$\chi $} is the
dielectric tensor; it is the sum of an identity tensor \underline{$I$} and
the electric susceptibility.  For ordinary nonabsorbing
materials the dielectric tensor is symmetric, so there always exists a
coordinate system with a set of axes, called the principal axes, in which
the dielectric tensor is diagonal. In optically uniaxial materials with the
optic axis in the $z$-direction, the $x$- and $y$-components of the
dielectric tensor are equal: 
\begin{equation}
\underline{\varepsilon }=\left( 
\begin{array}{ccc}
n_{0}^{2} & 0 & 0 \\ 
0 & n_{0}^{2} & 0 \\ 
0 & 0 & n_{e}^{2}%
\end{array}%
\right) \quad ,  \label{eq:eps}
\end{equation}%
where $n_{0}$ is called the ordinary and $n_{e}$ the extraordinary index of
refraction. We shall solve the wave equation only for a special case of the
wave vector direction, assuming that the wave vector is in the $xz$-plane: $%
\vec{k}=(k_{x},0,k_{z})$. For the electric field, we assume a general
direction $\overrightarrow{E}=(E_{x},E_{y},E_{z})$. One cannot assume that
the electric field is perpendicular to the wave vector, since such
orthogonality is valid only in isotropic materials. Using the ansatz for $%
\vec{k}$ and $\vec{E}$ and the expression (\ref{eq:eps}) for the dielectric
tensor, a set of three equations is obtained from the wave equation (Eq. (%
\ref{eq:valovnaen})): 
\begin{equation}
-k_{z}^{2}E_{x}+k_{x}k_{z}E_{z}+k_{0}^{2}n_{o}^{2}E_{x}=0\quad ,
\label{eq:we1}
\end{equation}%
\begin{equation}
-k_{z}^{2}E_{y}-k_{x}^{2}E_{y}+k_{0}^{2}n_{o}^{2}E_{y}=0\quad ,
\label{eq:we2}
\end{equation}%
\begin{equation}
-k_{x}^{2}E_{z}+k_{x}k_{z}E_{x}+k_{0}^{2}n_{e}^{2}E_{z}=0\quad .
\label{eq:we3}
\end{equation}%
The nontrivial solutions of the set of Eqs. (\ref{eq:we1}) - (\ref{eq:we3})
are obtained if the determinant of the coefficients in front of $E_{x,y,z}$
is zero. The determinant is zero if one of the conditions

\begin{equation}
k_{x}^{2}+k_{z}^{2}=k_{0}^{2}n_{o}^{2}  \label{eq:krog}
\end{equation}%
or

\begin{equation}
\frac{k_{x}^{2}}{n_{e}^{2}}+\frac{k_{z}^{2}}{n_{o}^{2}}=k_{0}^{2}
\label{eq:elipsa}
\end{equation}%
is satisfied.

The condition (\ref{eq:krog}) gives the magnitude of the wave vector in the
anisotropic material, a magnitute that is the same for all directions of
propagation in the $xz$-plane. This solution presents the ordinary wave. The
refractive index for the ordinary wave is isotropic and equal to $n_{o}$.
Eq. (\ref{eq:elipsa}) presents the constraint on the magnitude of the wave
vector of the extraordinary wave. The allowed magnitudes of the wave vectors
lie on an ellipse for the extraordinary wave and on a circle for the
ordinary wave (Fig. \ref{fig:elip}). When rotated around the optic axis, the
circle draws the surface of a sphere, and the ellipse the surface of an
ellipsoid; we thus obtain the wave vector surfaces, giving us the magnitude
of the wave vector for light propagation in a general direction. The reader
is referred to any book in modern optics to prove this, or he/she can check
it on his own by choosing different planes of wave vector direction (for
example, to the already studied $xz$-plane, consider also the $xy$- and the $%
yz$-plane).

Expressing the magnitude of the wave vector of the extraordinary wave as $%
k=k_{0}n(\theta )$, where $n(\theta )$ is the index of refraction of the
extraordinary wave for the light propagating with the wave vector at an
angle $\theta $ with respect to the optic axis, the wave vector can be
expressed as $\overrightarrow{k}=k(\sin \theta ,0,\cos \theta )$. Using this
expression, the angular dependence of the index of refraction can be derived
from Eq. (\ref{eq:elipsa}): 
\begin{equation}
n^{2}(\theta )=\frac{n_{e}^{2}n_{o}^{2}}{n_{o}^{2}\sin ^{2}\theta
+n_{e}^{2}\cos ^{2}\theta }\ \quad .  \label{eq:lomnikol}
\end{equation}%
The ordinary and extraordinary waves are linearly polarized. By using the
condition (\ref{eq:krog}) in Eqs. (\ref{eq:we1}) - (\ref{eq:we3}), we find
that $E_{x}=E_{z}=0$ and $E_{y}\neq 0$. The ordinary wave is thus linearly
polarized in the $y$-direction, i.e. in the direction perpendicular to the
plane defined by the optic axis and the wave vector. When the condition (\ref%
{eq:elipsa}) is used in Eqs. (\ref{eq:we1}) - (\ref{eq:we3}), we find that $%
E_{y}=0$ while $E_{x}$ and $E_{z}$ can be different from zero. The
polarization of the extraordinary wave is in the plane defined by the optic
axis and the wave vector and is thus perpendicular to the polarization of
the ordinary wave. From Eqs. (\ref{eq:we1}) and (\ref{eq:we3}), it is
straightforward to show that the polarization of the extraordinary wave is
along the $z$-axis if $\vec{k}$ is along the $x$-axis and vice versa. To
find the direction of $\vec{E}$ in the extraordinary wave with $\vec{k}$ at
a general angle with respect to the optic axis, is more elaborate. So we
state without a proof that $\vec{E}$ is always in the direction tangential
to the ellipse. The direction of the ray (meaning the direction of the
energy propagation and the direction of the ray velocity) at a given
direction of $\vec{k}$ (the direction of the phase velocity) is shown in
Fig. \ref{fig:elip}. The difference in the two directions is obvious.
However, one must bear in mind that even in materials with large anisotropy,
the ratio between the ordinary and the extraordinary index is of the order
of 10 \%, so the actual difference in the direction of propagation of $\vec{k%
}$ and $\vec{S}$ is much smaller. In calcite $n_{e}/n_{o}=1.486/1.658=0.896$%
, in quartz $n_{e}/n_{o}=1.553/1.544=1.006$ and in liquid crystals, where
the anisotropy is very large, $n_{e}/n_{o}=1.75/1.52=1.15$ for the liquid
crystal used in the experiment presented in this paper.

\subsection{Double refraction}

When a beam of unpolarized light is incident on the anisotropic uniaxial
transparent medium, it refracts into two beams. The direction of the wave
vectors for the two beams is determined by the boundary conditions at the
interface; these conditions follow from applying the Maxwell equations to
suitable regions containing the interface. The boundary conditions give us
the amplitude matching condition, which determines the reflectivity and the
transmissivity at the interface, and the phase matching condition, which
leads to the law of reflection and law of refraction. \cite%
{Rogalski,optics,Guenther,Fowles,Brooker} The phase matching condition
requires the components of the wave vector along the interface to be equal
in all the waves: the incident, the reflected and the refracted.

\begin{figure}[h]
\centering \includegraphics[scale=0.4]{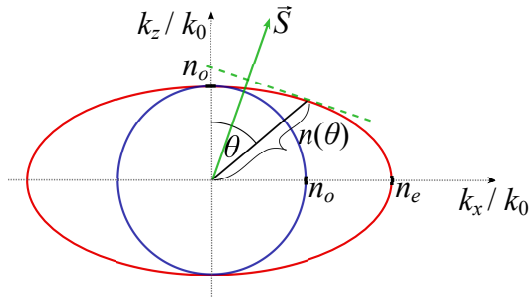}
\caption{\emph{(Color online)} The allowed wave vector magnitudes in the $xz$-plane for an
anisotropic material with the optic axis along the $z$-axis lie on the
circle (blue) for the ordinary wave and on the ellipse (red) for the
extraordinary wave. $\protect\theta $ is the angle between the wave vector $%
\vec{k}$ and the optic axis, $n_{o}$ and $n_{e}$ are the ordinary and the
extraordinary refractive index, respectively and $n$($\protect\theta $) is
the refractive index of the extraordinary wave, which propagates with $\vec{k%
}$ at an angle $\protect\theta $ with respect to the optic axis. At a given
direction of $\vec{k}$, the direction of the electric field $\vec{E}$ in the
extraordinary wave is tangential to the ellipse (green dashed line). The energy (defined by the
Poynting vector $\vec{S}=\vec{E}\times \vec{H}$, where $\vec{H}$ is the
magnetic field) propagates in the direction perpendicular to $\vec{E}$. $%
k_{x}$ and $k_{z}$ are the $x$- and $z$-component of the wave vector in the
anisotropic material and $k_{0}$ is the wave vector magnitude in vacuum.}
\label{fig:elip}
\end{figure}

\begin{figure}[h]
\centering \includegraphics[scale=0.6]{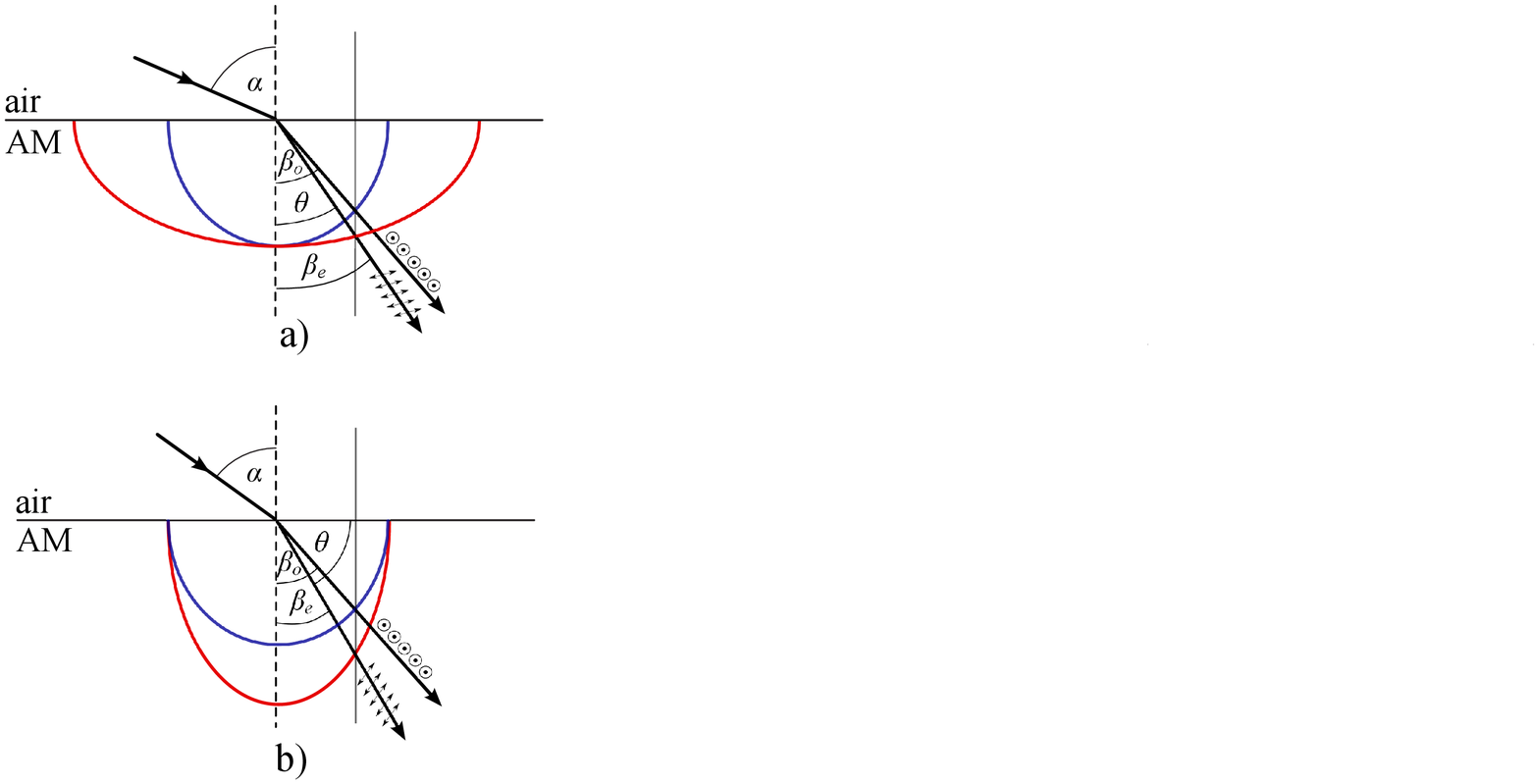}
\caption{\emph{(Color online)} Double refraction at the interface between the isotropic (air) and
anisotropic (AM) material if the optic axis lies in the plane of incidence
and is a) perpendicular and b) parallel to the interface. In the anisotropic
material the allowed magnitudes of the wave vector lie on the circle (blue)
for the ordinary \ wave and on the ellipse (red) for the extraordinary wave. 
$\protect\alpha $ is the incident angle, $\protect\beta _{o}$ and $\protect%
\beta _{e}$ are the refraction angles of the ordinary and extraordinary
wave, respectively, and $\protect\theta $ is the angle between the wave
vector of the extraordinary wave and the optic axis. The phase matching
boundary condition requires that the component of the wave vector parallel
to the interface be conserved. The magnitude of this component is presented
by a vertical solid line parallel to the interface normal (dashed line).
Polarization of the ordinary wave is in the direction perpendicular to the
incident plane (denoted by a circle with a dot), polarization of the
extraordinary wave is in the direction tangential to the ellipse and it is
not perpendicular to $\vec{k}$.}
\label{fig:front}
\end{figure}

Fig. \ref{fig:front} shows the refraction from an isotropic to an
anisotropic medium. In the anisotropic medium the wave vector surfaces are
shown for the ordinary and the extraordinary wave for the case of the optic
axis being perpendicular (Fig. \ref{fig:front}a) and parallel (Fig. \ref%
{fig:front}b) to the interface. The light is incident at an angle $\alpha $,
and $\beta _{o}$ and $\beta _{e}$ are the refraction angles for the wave
vector direction of the ordinary and the extraordinary wave, respectively.
Assuming that the isotropic medium is air with the refractive index
approximately 1, the phase matching boundary condition requires: 
\begin{equation}
k_{0}\sin \alpha =k_{o}\sin \beta _{o}\quad \mathrm{and}\quad k_{0}\sin
\alpha =k_{e}\sin \beta _{e}\quad ,  \label{eq:boundaryk}
\end{equation}%
where $k_{o}=k_{0}n_{o}$ and $k_{e}=k_{0}n(\theta )$ are the wave vector
magnitudes of the ordinary and the extraordinary wave, respectively. Eqs. (%
\ref{eq:boundaryk}) reduce to 
\begin{equation*}
\sin \alpha =n_{o}\sin \beta _{o}\quad \mathrm{and}\quad \sin \alpha
=n(\theta )\sin \beta _{e}\quad .
\end{equation*}%
We note that Snell's law is valid for the direction of the wave vector
propagation. However, since it is the direction of the energy propagation
(the ray direction) that we actually observe, Snell's law for the
extraordinary wave seems not to be obeyed (see Fig. \ref{fig:elip} and draw
the directions of energy propagation in Fig. \ref{fig:front}). The
refraction can be especially striking if the optic axis is at an angle with
respect to the interface, since one can observe the ray which refracts to
the same side of the normal (not only towards the normal). For the
demonstration of this fascinating property of anisotropic materials by using
liquid crystals, the reader is referred to Ref.~\onlinecite{oleg}.

\subsection{Optical anisotropy in liquid crystals}

Nematic liquid crystals are composed of elongated molecules with
orientationally ordered long molecular axes. The average direction of the
long molecular axes is called the director. Since all directions of
molecular motion in the direction perpendicular to the long molecular axis
are equally probable, the system is optically uniaxial with the optic axis
along the director. In bulky samples, clusters of oriented molecules are
formed, and the director varies in space. In nematic liquid crystals, the
orientational correlation length over which one can expect the same
orientation of molecules extends to 500 $\mu $m; therefore, the well-ordered
samples have to be thinner than that. \cite{fizikatk} Within the range of
optical frequencies, elongated molecules have greater polarizability along
the long molecular axis than perpendicular to it. Thus, when the external
electric field is along the director, induced polarization in the liquid
crystalline material will be larger than in the case when the electric field
is perpendicular to the director. Because of that, the light being polarized
perpendicularly to the director ($\vec{E}$ in the light is perpendicular to
the director) will be faster than the light with polarization parallel to
the director.

\section{Experiment}

\begin{figure}[h]
\centering \includegraphics[scale=0.7]{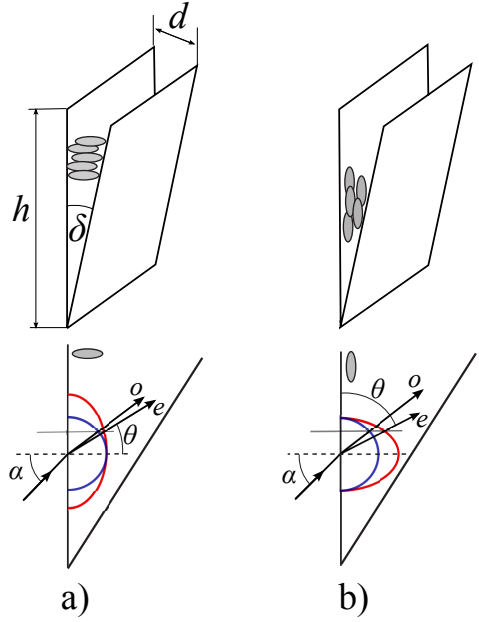}
\caption{\emph{(Color online)} The homeotropic and planar wedge cells. a) In the homeotropic cell
molecular long axes are oriented perpendicular to the cell surface and the
optic axis is thus perpendicular to the surface. The wedge angle $\protect%
\delta $ is defined by the length of the cell $h$ and the thickness of the
foil $d$. b) In the planar cell molecular long axes are oriented parallel to
the longer side of the cell. The optic axis is parallel to the surface. In
both cells the allowed magnitudes of the wave vectors of the ordinary (blue
circle) and the extraordinary (red ellipse) wave are shown (compare also to
Fig. 2). The wave vector directions of the ordinary (\textit{o}) and the
extraordinary (\textit{e}) waves are shown. $\protect\alpha $ is the
incident angle and $\protect\theta $ is the angle between the wave vector of
the extraordinary wave and the optic axis.}
\label{fig:celice}
\end{figure}

\begin{figure}[h]
\centering \includegraphics[scale=0.6]{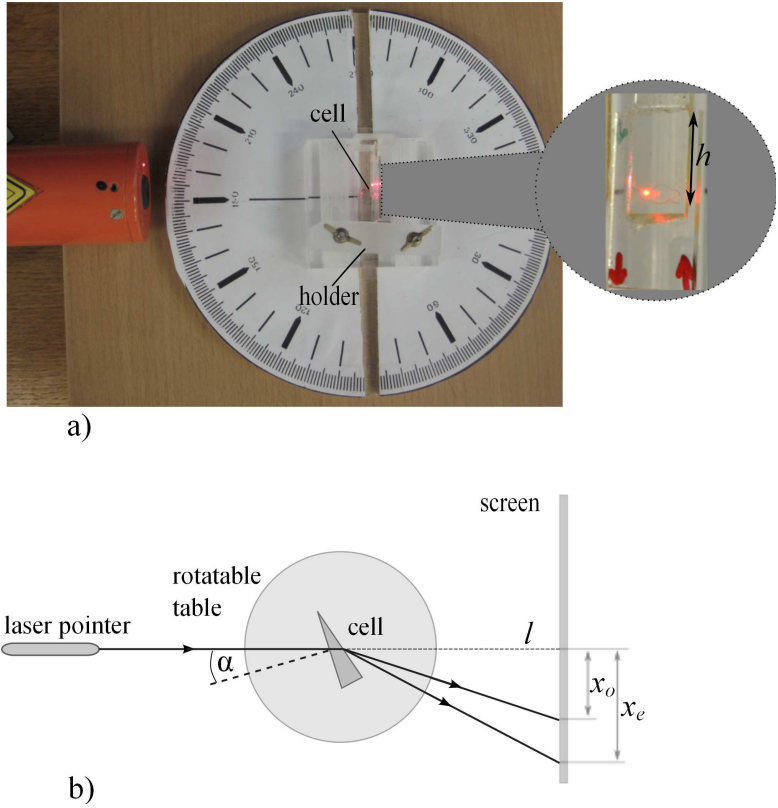}
\caption{a) \emph{(Color online)} Experimental setup with a zoomed liquid
crystalline cell. b) Schematic presentation. $l$ is the distance between the
cell and the screen. $x_{o}$ and $x_{e}$ define the position of the ordinary
and extraordinary beam, respectively. For example, if one takes a planar
cell with a wedge angle $\protect\delta =3.2^{\circ }$, then at the incident
angle $\protect\alpha =10^{\circ }$ and $l=5.07$ m, the ordinary and the
extraordinary beams hit the screen at $x_{o}=15.5$ cm and $x_{e}=22.0$ cm,
respectively.}
\label{fig:postavitev}
\end{figure}

To provide the refraction situations shown in Fig. \ref{fig:front}, two
wedge cells were designed with different surface treatments in order to
achieve different orientations of the director in the cell and thus
different orientations of the optic axis with respect to the surface (Fig. %
\ref{fig:celice}). The cells were approximately 1 cm long ($h$) and half a
centimeter wide (Fig. \ref{fig:postavitev}a). Usual laser beams are too wide
to enable studies of double refraction in cells having parallel surfaces. In
thin samples, which guarantee homogeneity of orientation of the long
molecular axis (the director gives the direction of the optical axis), the
spatial separation of the ordinary and extraordinary beam can be obtained by
the prismatic effect. \cite{klin, nclc} To prepare a wedge cell, a foil with
a thickness $d=360\ \mu \mathrm{m}$ was inserted and glued between two
pieces of microscope glass in one of the narrower sides, while the other
narrow side of the cell was glued together directly, thus forming a wedge.
By rubbing the surfaces, the planar cell in which molecules are aligned in
the surface plane along the long side of the surface (Fig. \ref{fig:celice}%
b) was designed. The elongated molecules align with their long axes along
the scratches, and the director is parallel to the rubbing direction.
Therefore, the optic axis also coincides with the rubbing direction. To
align the molecules perpendicular to the surface (homeotropic cell, Fig. \ref%
{fig:celice}a), a polymer coating was applied to the glass. Professional
cells usually use carefully engineered coatings, but for simple experiments
a satisfactory effect is provided by simply dipping the glass into a
detergent or lecithin solution and allowing it to dry. The capillary effect
was used to fill the cells with the liquid crystal E18 heated above the
transition temperature from the nematic to the isotropic phase. \cite{where
to buy} For E18 at room temperature, the values of the ordinary and
extraordinary indices are $n_{0}=1.52$ and $n_{e}=1.75$. \cite{E18}

The experimental setup is shown in Fig. \ref{fig:postavitev}a. The wedge
cell is fixed into a holder and placed on a rotatable table with the longer
side parallel to the table surface. A\ He-Ne laser is used as the source for
the unpolarized light. The direction of the incident light is always in
the plane perpendicular to the wedge. When the laser beam of unpolarized
light is incident on the wedge cell, two bright spots are observed on a
distant screen; this is because of the birefringence of the liquid crystal
in the wedge cell. The refractive indices are determined by measurements of
the relative position of the spots with respect to the position of the
direct beam spot, when light does not pass through the cell (Fig. \ref%
{fig:postavitev}b). The position of spots changes as the incident angle of
light alters when the table rotates; this enables measurements of the
angular dependence of both indices.

\begin{figure}[h]
\centering \includegraphics[scale=0.55]{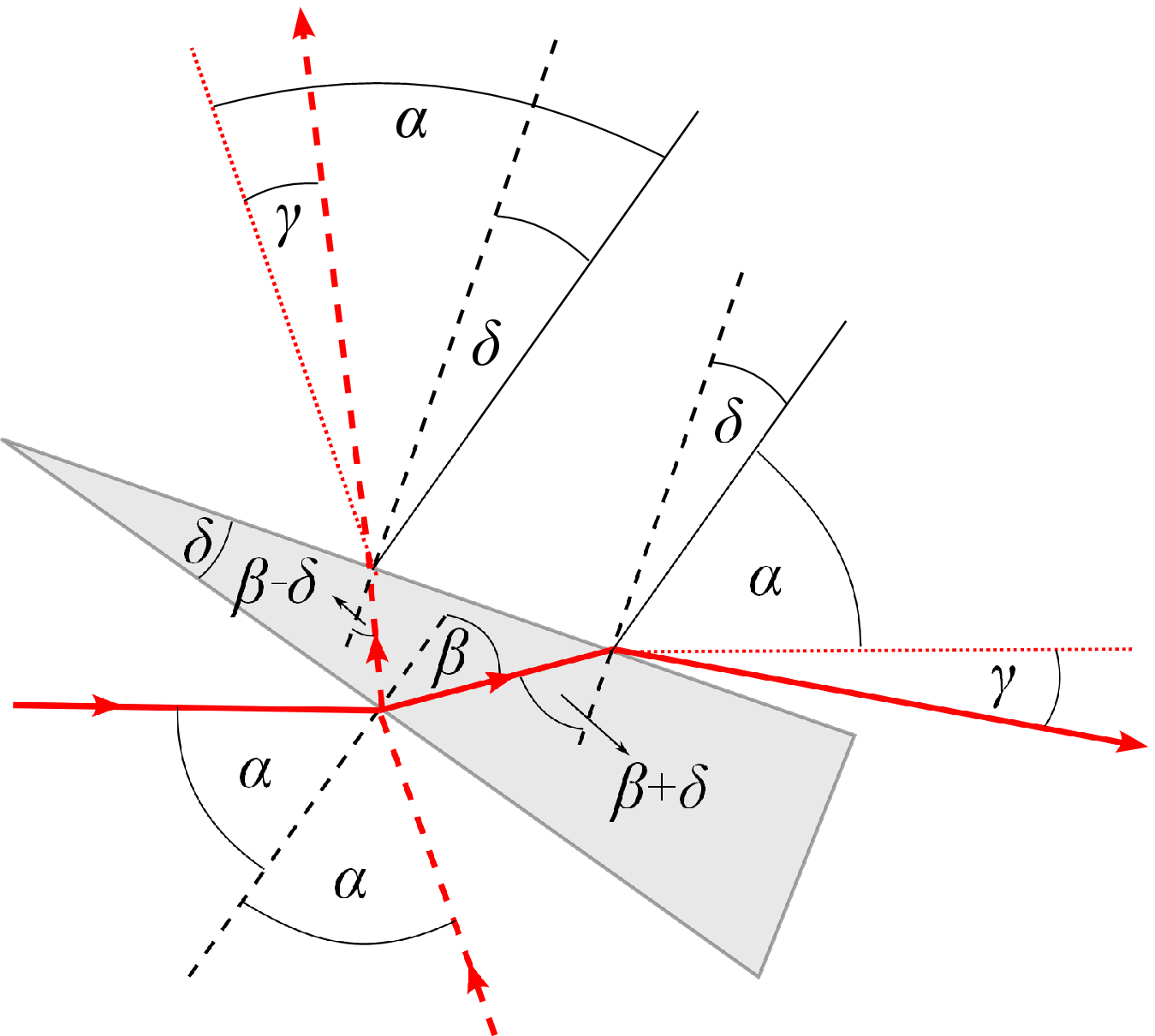}
\caption{The geometry of the experiment. Angles are defined in the text. }
\label{fig:koti}
\end{figure}

Fig. \ref{fig:koti} shows the geometry of light propagation through the
wedge cell. The angle $\alpha $ is the controlled incident angle, and $\beta 
$ is the refraction angle giving the direction of the wave vector of either
the ordinary ($\beta _{o}$) or the extraordinary ($\beta _{e}$) wave. When
exiting the cell the direction of light propagation deviates from the
direction of the incident beam by an angle of $\gamma $, which differs for
the ordinary and the extraordinary wave and can be calculated from the beam
position on the screen and the distance between the cell and the screen as $%
\gamma \approx x/l$ (see Fig. \ref{fig:postavitev}b) where, again, one
should use $x$ for either the ordinary ($x_{o}$) or the extraordinary ($x_{e}
$) wave.

The refractive indices of the ordinary and the extraordinary wave are
obtained by measuring $\gamma $ as a function of $\alpha $ and knowing the
wedge angle of the cell $\delta \approx d/h$ (Fig. \ref{fig:celice}).
Applying Snell's law at the two interfaces of the wedge cell, we find (see
Fig. \ref{fig:koti}): 
\begin{equation}
{\frac{\sin \alpha }{\sin \beta }}=n\quad \mathrm{and}\quad {\frac{\sin
(\alpha \pm \gamma \pm \delta )}{\sin (\beta \pm \delta )}}=n\quad ,
\label{eq:sinusi}
\end{equation}%
where $n$ is the refractive index of either the ordinary or the
extraordinary wave. In Eq. (\ref{eq:sinusi}) the upper sign in $\pm $ stands
for the beam direction given in Fig. \ref{fig:koti} by the red solid line
(we shall call this incident angle positive $\alpha $), and the lower sign
for the beam denoted by the red dashed line (negative $\alpha $). Eq. (\ref%
{eq:sinusi}) is an approximation, since the direction of the optic axis in
the cell varies slightly because of the wedge. However, since $\delta $ is
small, one can assume that the refractive indices of the extraordinary wave  at angles $%
\beta $ and $\beta \pm \delta $ are the same. To confirm this, we measured $%
x_{e}$ at incidence angles $\pm \alpha $ and found them the same within the
experimental error and the width of the beam spots on the screen.

One should also note that, since in the extraordinary wave the energy and
the wave vector do not propagate in the same direction in the liquid crystal
cell, the extraordinary ray exiting the cell will be displaced from the
position shown in Fig. \ref{fig:koti}, where the wave vector directions are
drawn. However, the displacement is of the order of cell thickness and can
thus be neglected.

Since the wedge angle $\delta $ and the angle $\gamma $ are very small, $%
\sin \delta $ $\approx $ $\delta $, $\cos \delta \approx 1$ and $\sin \gamma
\approx \gamma $, $\cos \gamma \approx 1$. By using the sine addition
formulae and equating the left parts of Eqs. (\ref{eq:sinusi}), we find (up
to the first order in $\delta $ and $\gamma $):

\begin{equation}
{\frac{\sin \alpha }{\sin \beta }}={\frac{\sin \alpha +\gamma \cos \alpha
+\delta \cos \alpha }{\sin \beta +\delta \cos \beta }}\quad 
\label{eq:poenostavitev}
\end{equation}%
from where the refraction angle $\beta $ follows: 
\begin{equation}
\tan \beta ={\frac{\delta }{\delta +\gamma }}\tan \alpha \quad .
\label{eq:tangens}
\end{equation}%
With $\beta $ obtained from Eq. (\ref{eq:tangens}), the value of the
refractive index is found from Eq. (\ref{eq:sinusi}). Eqs. (\ref{eq:sinusi})
and (\ref{eq:tangens}) are general expressions that can be used to obtain
the refractive index when light passes through a thin wedge sample of any,
not necessarily birefringent, material and they allow evaluation of the
refractive indices of both the ordinary and the extraordinary wave exiting a
cell filled by birefringent material.

In the homeotropic cell (Fig. \ref{fig:celice} a), the optic axis is
perpendicular to the glass plate, and the refraction angle of the
extraordinary wave ($\beta _{e}$) is equal to the angle $\theta $ between
the optic axis and the wave vector, so the refractive index of the
extraordinary wave is:

\begin{equation}
n(\theta )={\frac{\sin \alpha }{\sin \beta _{e}}}\quad ,
\label{homeotropnasin}
\end{equation}%
where $\beta _{e}$ is calculated from Eq. (\ref{eq:tangens}) using the
measured value of $\gamma $ for the extraordinary wave.

From the measurements performed on the homeotropic cell, the refractive
indices of the ordinary and extraordinary wave at angles ranging from $%
\theta =0^{\circ }$ to approximately $40^{\circ }$ can be obtained.

In the planar cell the optic axis is parallel to the glass, and the
refraction angle of the extraordinary wave is related to the angle between
the optic axis and the wave vector direction as $\beta _{e}={\frac{\pi }{2}}%
-\theta $. The refractive index of the extraordinary wave is therefore given
by:

\begin{equation}
{\frac{\sin \alpha }{\sin \beta _{e}}}={\frac{\sin \alpha }{{\sin {({\frac{%
\pi }{2}}}-\theta )}}}=n(\theta )\quad .  \label{planarnasin}
\end{equation}%
From the measurements in the planar cell, the values are obtained for the
refractive index at angles $\theta $ ranging from approximately $60^{\circ }$
to $90^{\circ }$ . The planar cell can thus be used to study the direction
dependency of the refractive index of the extraordinary wave in the region
where the difference in the indices of the ordinary and the extraordinary
wave is close to its largest value.

\begin{figure}[h]
\centering \includegraphics[scale=0.55]{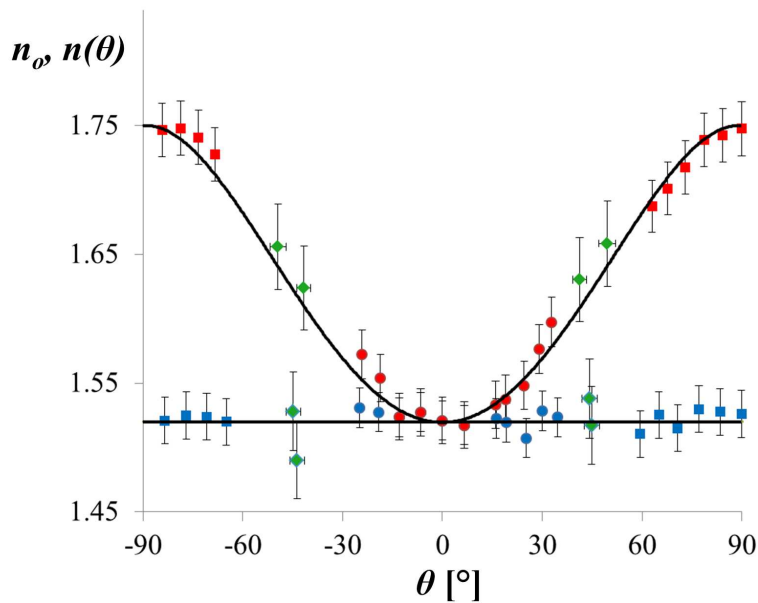}
\caption{\emph{(Color online)} The refractive indices of the ordinary ($n_{o}
$) and the extraordinary ($n(\protect\theta )$) wave as a function of the
angle $\protect\theta $ between the optic axis and the wave vector direction
in the nematic liquid crystal E18. Squares: values obtained from the
measurements in the planar cell. Circles: values obtained from the
measurements in the homeotropic cell. Red: $n(\protect\theta )$, Blue: $n_{o}
$. Green diamonds: measurements of $n_{o}$ and $n(\protect\theta )$ in the
sandwiched cell (see Fig. 8). The theoretical dependence $n(\protect\theta )$
given by Eq. (\protect\ref{eq:lomnikol}) with $n_{e}=1.75$ and $n_{o}=1.52$
is given for comparison.}
\label{fig:graf}
\end{figure}

There are a few limitations in the experiment that must be considered.
Although the ordinary and extraordinary wave can propagate in any direction,
experimentally we are limited with the refraction of the incident light,
since the light source is outside the birefringent material. In the
presented situation, the wave vector direction of the ordinary wave was
theoretically limited (at the incident angle $\alpha =90^{\circ }$) to $%
\theta =41^{\circ }$ at $n_{o}=1.52$ and for the extraordinary wave to $%
\theta =35^{\circ }$ at $n_{e}=1.75$. The cell size and the cell holder
additionally limit the incident angle $\alpha $. Fig. \ref{fig:graf} shows
the combined results of the angular dependence of the refractive indices of
the ordinary and the extraordinary wave measured in the planar and the
homeotropic cell. The theoretical dependence was calculated from Eq. (\ref%
{eq:lomnikol}) using the known values of $n_{o}$ and $n_{e}$.

From Fig. \ref{fig:graf} it can clearly be seen that with such a simple
experiment we were not able to measure the values of the refractive indices
for directions of propagation close to $\theta =45^{\circ }$. To measure the
refractive indices for these directions as well, one should use an old
experimental trick: phase matching with the help of an additional material. 
\cite{trik} In order to enter the liquid crystal under a general angle, one
should prevent refraction between two materials that have a large difference
in refractive indices. Therefore, the light should pass a surface where
refractive indices of materials on both sides of the surface are similar.

\begin{figure}[h]
\centering \includegraphics[scale=0.5]{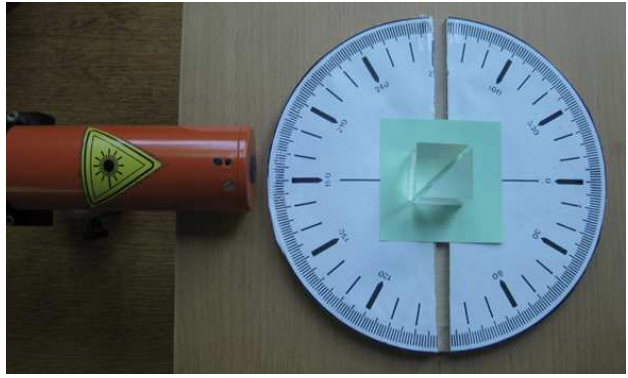}
\caption{\emph{(Color online)} The sandwich of prisms and the cell.}
\label{fig:slikasendvica}
\end{figure}

To achieve such conditions, the wedge cell is sandwiched between two glass
prisms (refractive index $n_{g}$ = 1.50) with an apex angle of $45^{\circ }$
(Fig. \ref{fig:slikasendvica}). To prevent any possibility of a tiny air
interface between the prism and the wedge cell, the contact areas are
covered by glycerol, which has a refractive index similar to the refractive
indices of glass and liquid crystal. \cite{glycerol} If the incident angle
on the prism surface is zero, the beam does not refract and the incident
angle to the liquid crystal in the wedge is $45^{\circ \text{ }}$(Fig. \ref%
{fig:sendvicshema}). Since the refractive indices of glass and liquid
crystal are similar, the angle of refraction does not differ much from the
incident angle, and in the liquid crystal light propagates in a direction
close to $\theta =45^{\circ }$. The prism at the other side of the cell
provides a change in light direction that is opposite to the first one.
Without the wedge cell, the light beam should be straight. Therefore the
positions of the two light spots again allow simultaneous measurement of
both refractive indices.

\begin{figure}[h]
\centering \includegraphics[scale=0.6]{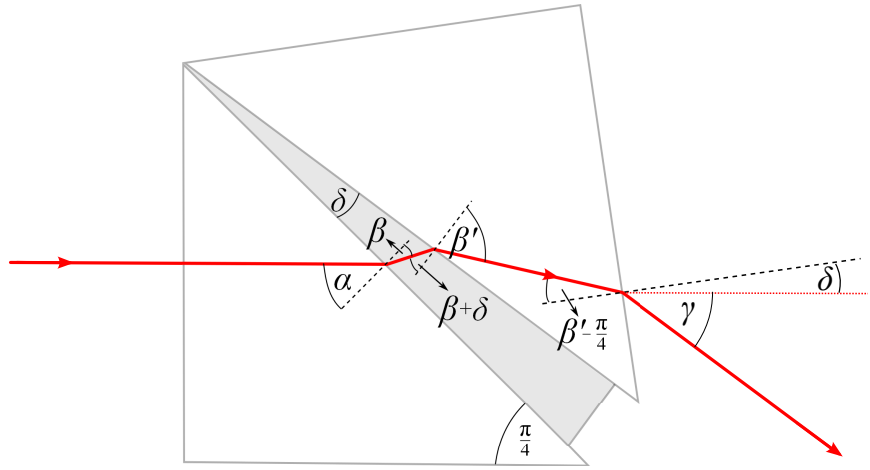}
\caption{The geometry of the upgraded experiment (sandwich). The angles are
defined in the text.}
\label{fig:sendvicshema}
\end{figure}

Let us follow the light beam through the sandwich and calculate the
refractive indices (Fig. \ref{fig:sendvicshema}). If the glycerol film is of
uniform thickness, it does not influence the analysis of the refractive
indices. The first relevant interface is thus the glass - liquid crystal
interface. The incident angle is $\alpha =45^{\circ }$, and the beam
refracts only slightly ($\beta $ is close to $45^{\circ }$) because the
refractive indices of liquid crystal are close to 1.5. At the liquid crystal
- glass interface, the incident angle changes, owing to the wedge angle $%
\delta $, to $\beta +\delta $, as in Fig. \ref{fig:sendvicshema}, or to $%
\beta -\delta $ if the sandwich is rotated $90^{\circ }$ clockwise.
Therefore, Snell's law gives:

\begin{equation}
{\frac{\sin (\beta \pm \delta )}{\sin \beta ^{\prime }}}={\frac{n_{g}}{n}%
\quad ,}  \label{razmerjen}
\end{equation}%
where $\beta ^{\prime }$ is the refraction angle from the liquid crystal to
the glass (see Fig. \ref{fig:sendvicshema}). The last refraction occurs at
the glass - air interface. The incident angle is equal to $\beta ^{\prime
}-\pi /4$, as in Fig. \ref{fig:sendvicshema}, or $\beta ^{\prime }+\pi /4$
if the sandwich is rotated 90 degrees clockwise, as in Fig. \ref%
{fig:slikasendvica}. One can write Snell's law as

\begin{equation}
\sin (\beta ^{\prime }\mp \pi /4)={\frac{\sin (\gamma \pm \delta )}{n_{g}}%
\quad .}  \label{betadvecrti}
\end{equation}

The angle $\theta $ between the optic axis and the wave vector direction of
the extraordinary/ordinary wave is equal to $\beta $ in the homeotropic cell
and $\pi /2-\theta $ in the planar cell (see Fig. \ref{fig:celice}).

As explained above, the sandwiched cells provide extra measurements of
refractive indices at angles $\theta $ close to $45^{\circ }$. Since we have
two cells with different alignments of molecules and two different
orientations of sandwich as in Figs. \ref{fig:slikasendvica} and \ref%
{fig:sendvicshema}, the refractive indices in four different directions of
light propagation in the cell with respect to the optical axis can be
measured. These measurements are shown as green diamonds in Fig. \ref%
{fig:graf}. One can clearly see that the accuracy of the refractive index
measurement when the wedge cell is fixed between the prisms is much lower
than when the prisms are absent. The reason is the glycerol, because it
forms a slight wedge as well, which could not be precisely controlled. In
most cases, the wedge of the glycerol has the same orientation as the wedge
of the liquid crystalline cell, a fact which results in larger values for
refractive indices. Less frequently, the glycerol wedge is in the opposite
direction, which leads to lower values for the refractive index. The effect
was verified in the absence of the liquid crystal. Nevertheless, the
experiment nicely shows that values of the refractive indices at $\theta $\
between $40^{\circ }$ and $50^{\circ }$ are consistent with the calculated
direction dependence of the refractive index of the extraordinary wave in
the uniaxial liquid crystal.

\section{Conclusion}

Students are confronted by the difficult concept of birefringence during
physics lessons at university. We have shown that the concept can be fully
demonstrated by using nematic liquid crystalline wedge cells with different
orientation of molecules. Most important, the experiment, the setup of which
consists of a laser, rotatable table, two different wedge cells, two prisms
with an apex angle of $45^{\circ }$ and a drop of glycerol, enables
quantitative measurements of the angular dependence of the refractive index
of the extraordinary wave for the whole range of propagation directions. The
homeotropic cell permits the demonstration of the direction dependency of
the refractive index of the extraordinary wave, as well as its measurement,
when its value is close to the value of the ordinary refractive index. The
planar cell can be used for quantitative measurement of the refractive index
direction dependency close to its largest value. By using additional glass
prisms to form a sandwich of prisms and the wedge cell, one can study the
angular dependency of the refractive indices in a range of directions that
cannot be attained with a simple setup.

\begin{acknowledgments}
The authors are grateful to Miha \v{S}karabot for
producing the wedge cells, to Martina \v{S}ubic for the control set of
measurements and to Igor Mu\v{s}evi\v{c} for pointing out the paper by Barnik \textit{et al}. ~\cite{trik}
This work was supported by the project J5-4002 financed by the Slovenian Research Agency (ARRS).
\end{acknowledgments}


\begin{thebibliography}{99}
\bibitem{nclc} P. Oswald, and P. Pieranski, \textit{Nematic and Cholesteric
Liquid Crystals: Concepts and Physical Properties}, (Taylor and Francis, New
York, 2005).

\bibitem{knitting} M. \v{C}epi\v{c}, \textquotedblleft Knitted patterns as a
model for anisotropy,\textquotedblright\ Phys. Educ. \textbf{47}, 456-61,
(2012).

\bibitem{ciferno} T. M. Ciferno, R. J. Ondris-Crawford, and G. P. Crawford,
\textquotedblleft Inexpensive electrooptic experiments on liquid crystal
displays,\textquotedblright\ Phys. Teach. \textbf{33}(2), 104-10 (1995).

\bibitem{Pavlin} J. Pavlin, K. Susman, S. Ziherl, N. Vaupoti\v{c}, and M. 
\v{C}epi\v{c}, \textquotedblleft How to teach liquid
crystals,\textquotedblright\ Mol. Cryst. Liq. Cryst. \textbf{547}, 255-61
(2011).

\bibitem{pieranski} P. Pieranski, \textquotedblleft Classroom experiments
with chiral liquid crystals,\textquotedblright\ in \textit{Chirality in
Liquid Crystals}, edited by H. Kitzerow and Ch. Bahr, (Springer, New York,
2001).

\bibitem{calcite} E. Hecht, \textit{Optics}, 3rd ed. (Addison Wesley
Longman, Reading, 1998).

\bibitem{Verbit} L. Verbit, \textquotedblleft Liquid Crystals: Synthesis and
Properties,\textquotedblright\ J. Chem. Educ., \textbf{49}(1), 36-39 (1972).

\bibitem{Wright} J. J. Wright, \textquotedblleft Optics Experiments with
Nematic Liquid Crystals,\textquotedblright\ Am. J. Phys., \textbf{41} 2),
270-72 (1973).

\bibitem{Liberko} C. A. Liberko, and J. Shearer, \textquotedblleft
Preparation of a surface-oriented liquid crystal - An experiment for the
undergraduate organic chemistry laboratory,\textquotedblright\ J. Chem.
Educ. \textbf{77}(9), 1204-05 (2000).

\bibitem{Eurasia} J. Pavlin, N. Vaupoti\v{c}, S. A. Gla\v{z}ar, M. \v{C}epi%
\v{c}, and I. Devetak, \textquotedblleft Slovenian pre-service teachers'
conceptions about liquid crystals,\textquotedblright\ Eurasia, \textbf{7}%
(3), 173-80 (2011).

\bibitem{nasvet-valen} For a quick overview, we suggest the concise
treatment in Ref.~\onlinecite{Rogalski}, but, of course, any book on optics
or modern optics (as Refs.~\onlinecite{optics,Guenther,Fowles,Brooker}) will
also have a thorough treatment of light propagation in anisotropic media.

\bibitem{Rogalski} M. S. Rogalski and S. B. Palmer, \textit{Advanced
University Physics}, 2nd ed. (Chapman \& Hall/CRC Taylor, Francis Group,
Boca Raton, 2006).

\bibitem{optics} M. Born and E. Wolf, \textit{Principles of Optics}, 7th
(expanded) ed. (Cambridge University Press, Cambridge, 1999).

\bibitem{Guenther} R. Guenther, \textit{Modern Optics} (John Wiley \& Sons,
New York, 1990).

\bibitem{Fowles} G. R. Fowles, \textit{Modern Optics}, 2nd ed. (Dover
Publications, New York, 1989).

\bibitem{Brooker} G. Brooker, \textit{Modern Classical Optics} (Oxford
University Press, Oxford, 2011).

\bibitem{oleg} O. P. Pishnyak, and O. D. Lavrentovich, \textquotedblleft
Electrically controlled negative refraction in a nematic liquid
crystal,\textquotedblright\ Appl. Phys. Lett. \textbf{89}, 251103 (2006).

\bibitem{fizikatk} P. G. de Gennes, and J. Prost, \textit{The Physics of
Liquid Crystals}, 2nd ed. (Oxford Science Publications, Oxford, 1995).

\bibitem{klin} D. K. Shenoy, \textquotedblleft Measurements of Liquid
Crystal Refractive Indices,\textquotedblright\ Am. J. Phys. \textbf{62},
858-59 (1994).

\bibitem{where to buy} Nematic liquid crystals with appropriate properties
are commercially available from several companies like Merck, Sigma Aldrich
or Nematel. Different commercial mixtures are available which are in the
liquid crystalline state around the room temperature. The measurements can
be done with any nematic liquid crystals that have the nematic phase at the
room temperature and have high birefringence. Instead, we can also use the
liquid crystal MBBA which is quite easy to synthesize in the school lab. We have used E18, which was available in the lab, however, this
mixture cannot be bought any more.

\bibitem{E18} BDH Catalog, BDH Chemical Ltd., Poole BH 12 4NN,
England 1978, p. 21.

\bibitem{glycerol} Refractive index of glycerol, $\langle$%
http://refractiveindex.info/$\rangle$.

\bibitem{trik} M. I. Barnik, L. M. Blinov, A. M. Dorozhkin, and N. M.
Shtykov, \textquotedblleft Generation of the second optical harmonic induced
by an electric field in nematic and smectic liquid
crystals,\textquotedblright\ Sov. Phys. JETP \textbf{54}(5), 935-38 (1981).
\end{thebibliography}
\end{document}